\documentclass[11pt]{article}

\usepackage{amsmath,amssymb,amsfonts,amsthm,amstext}
\usepackage{thmtools}
\usepackage{mathtools}
\usepackage{fullpage}
\usepackage{tikz}

\theoremstyle{definition}
\newtheorem{problem}{Problem}
\newtheorem{fact}{Fact}
\newtheorem{algorithm}{Algorithm}
\newtheorem{definition}{Definition}
\newtheorem{example}{Example}
\newtheorem{theorem}{Theorem}
\newtheorem{lemma}{Lemma}
\newtheorem{remark}{Remark}
\newtheorem{corollary}{Corollary}

\newcommand{\ket}[1]{|#1\rangle}
\newcommand{\bra}[1]{\langle#1|}

\newcommand{\C}{\mathbb{C}}

\newcommand{\N}{\mathbb{N}}
\newcommand{\F}{\mathbb{F}}
\newcommand{\Z}{\mathbb{Z}}

\newcommand{\QFT}{\mathrm{QFT}}

\begin{document}

\title{Quantum algorithms for abelian difference sets\\
and applications to dihedral hidden subgroups}

\author{Martin Roetteler\\[2ex]
Microsoft Research\\
Quantum Architectures and Computation Group\\
One Microsoft Way, Redmond, WA 98052, U.S.A.\\
\texttt{martinro@microsoft.com}
}

\maketitle

\begin{abstract}
Difference sets are basic combinatorial structures that have applications in signal processing, coding theory, and cryptography. We consider the problem of identifying a shifted version of the characteristic function of a (known) difference set. We present a generic quantum algorithm that can be used to tackle any hidden shift problem for any difference set in any abelian group. We discuss special cases of this framework where the resulting quantum algorithm is efficient. This includes: a) Paley difference sets based on quadratic residues in finite fields, which allows to recover the shifted Legendre function quantum algorithm, b) Hadamard difference sets, which allows to recover the shifted bent function quantum algorithm, and c) Singer difference sets based on finite geometries. The latter class allows us to define instances of the dihedral hidden subgroup problem that can be efficiently solved on a quantum computer. 
\end{abstract}

%
%

\section{Introduction} \label{sect:Intro}

Many exponential speedups in quantum computing are the result of solving problems that belong to either the class of hidden subgroup problems (HSPs) or the class of hidden shift problems. For instance, the problems of factoring integers and of computing discrete logarithms in abelian groups \cite{Factoring} can be reformulated as solving instances of hidden subgroup problems in abelian groups \cite{MoscaEkert,Kitaev95,Kitaev-book,BH:97,Jozsa98,Jozsa01}: given a function $f$ from an abelian group $A$ to a set, so that $f$ is constant on the cosets of some subgroup $H\leq A$ and takes distinct values on different cosets, the task is to find generators of $H$. 

Successes of the hidden subgroup framework include period finding over the reals which was used by Hallgren to construct an efficient quantum algorithm for solving Pell's equation~\cite{Hallgren, JozsaPell} and more recently to the discovery of a quantum algorithm for computing unit groups of number fields of arbitrary degree \cite{EHKS:2014}. Moreover, the hidden subgroup problem over symmetric and dihedral groups are related to the graph isomorphism problem~\cite{BonehLipton, Beals, EttingerHoyer, HMRRS:2010} and some computational lattice problems~\cite{Regev:2004}. Constructing efficient algorithms for these problems are two major open questions in quantum algorithms. 

As far as hidden shift problems are concerned, the shifted Legendre function problem \cite{vDHI:2003}, shifted sphere problems and shifts of other non-linear structures \cite{CSV:2007}, problems of finding shifts of non-linear Boolean functions \cite{Roetteler:2010,CKOR:2013} can be reformulated as solving instances of hidden shift problems in abelian groups: given a pair $(f,g)$ of functions from an abelian group $A$ to a set so that $g$ is obtained from $f$ by shifting the argument by an unknown shift $s\in A$, the task is to find this shift. For further background on hidden subgroup and hidden shift problems see~\cite{NC, Kitaev-book, KLM, CvD10, Lomont}.

An intriguing connection exists between {\em injective} instances of the hidden shift problem over abelian groups $A$ and the hidden subgroup problem for semidirect products of the form $A \rtimes \Z_2$ where the action of $\Z_2$ is given by inversion. This connection includes the special case of the hidden shift problem over the cyclic groups $A=\Z_N$, where $N$ is a large integer which are related to the dihedral groups $D_N = \Z_N \rtimes \Z_2$. Despite much effort, a fully polynomial-time quantum algorithm for the hidden subgroup problem over the dihedral groups has remained elusive. In this paper we make a step toward solving the hidden subgroup problem over the dihedral groups by exhibiting some instances that can be solved efficiently on a quantum computer. By efficient we mean that the run-time of the quantum part of the computation is bounded polynomially in the input size, which is generally assumed to be $\log{A}$, and the run-time of the classical post-processing part of the computation is also bounded polynomially in the input size. 
\begin{figure}[hbt]
\centerline{
\begin{tabular}{cccc}
\raisebox{2mm}{\includegraphics[width=8cm]{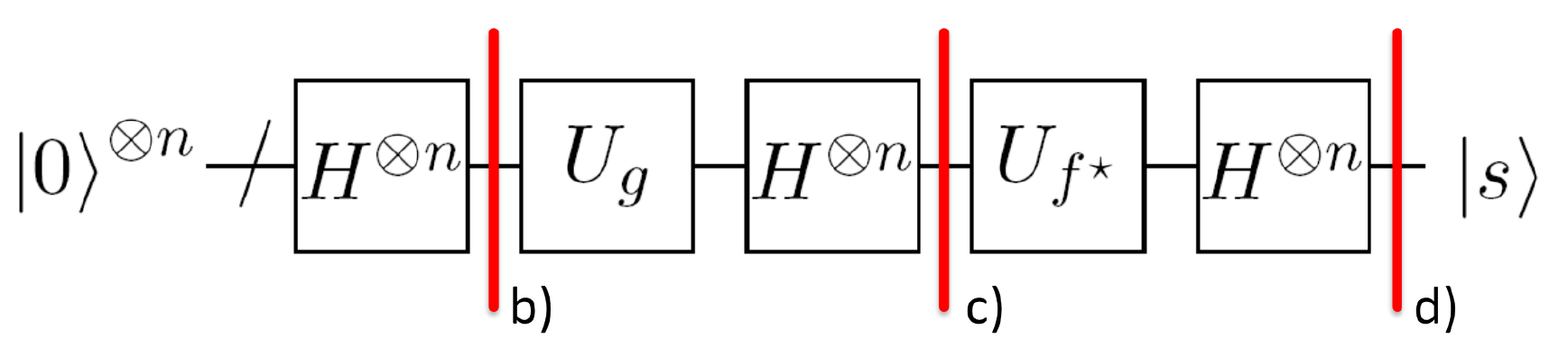}} &
\includegraphics[width=2.3cm]{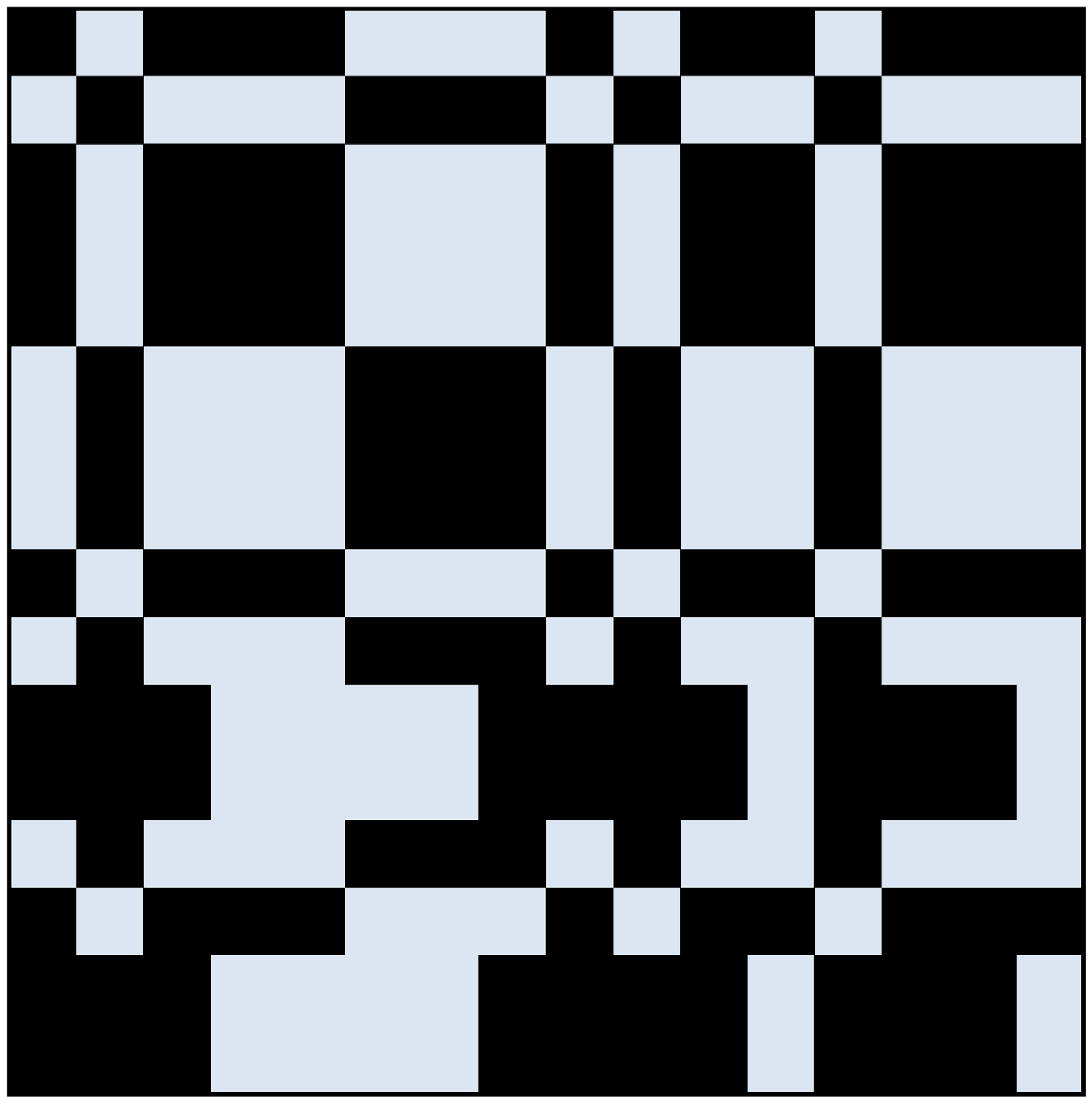} & 
\includegraphics[width=2.3cm]{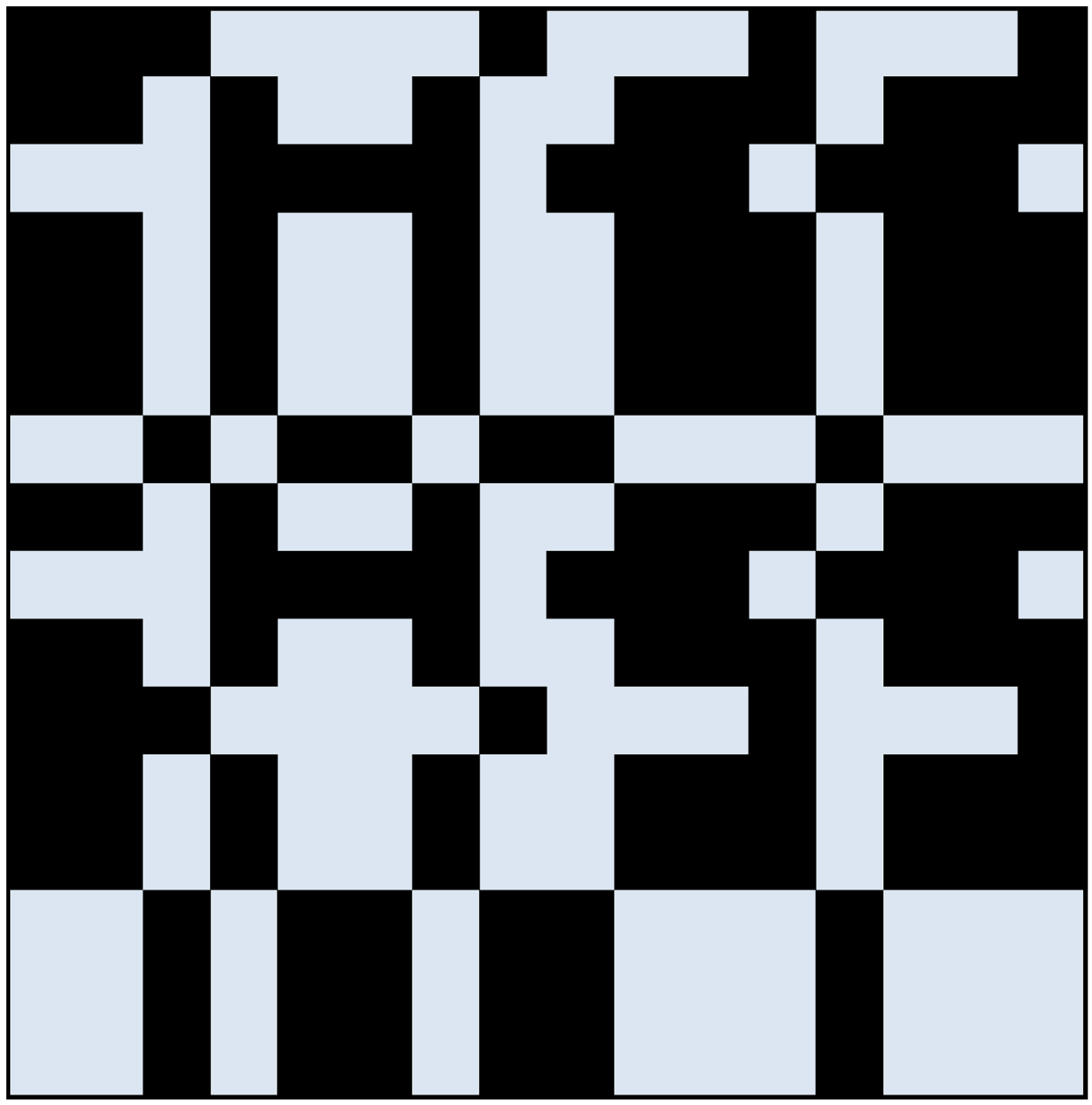} & 
\includegraphics[width=2.3cm]{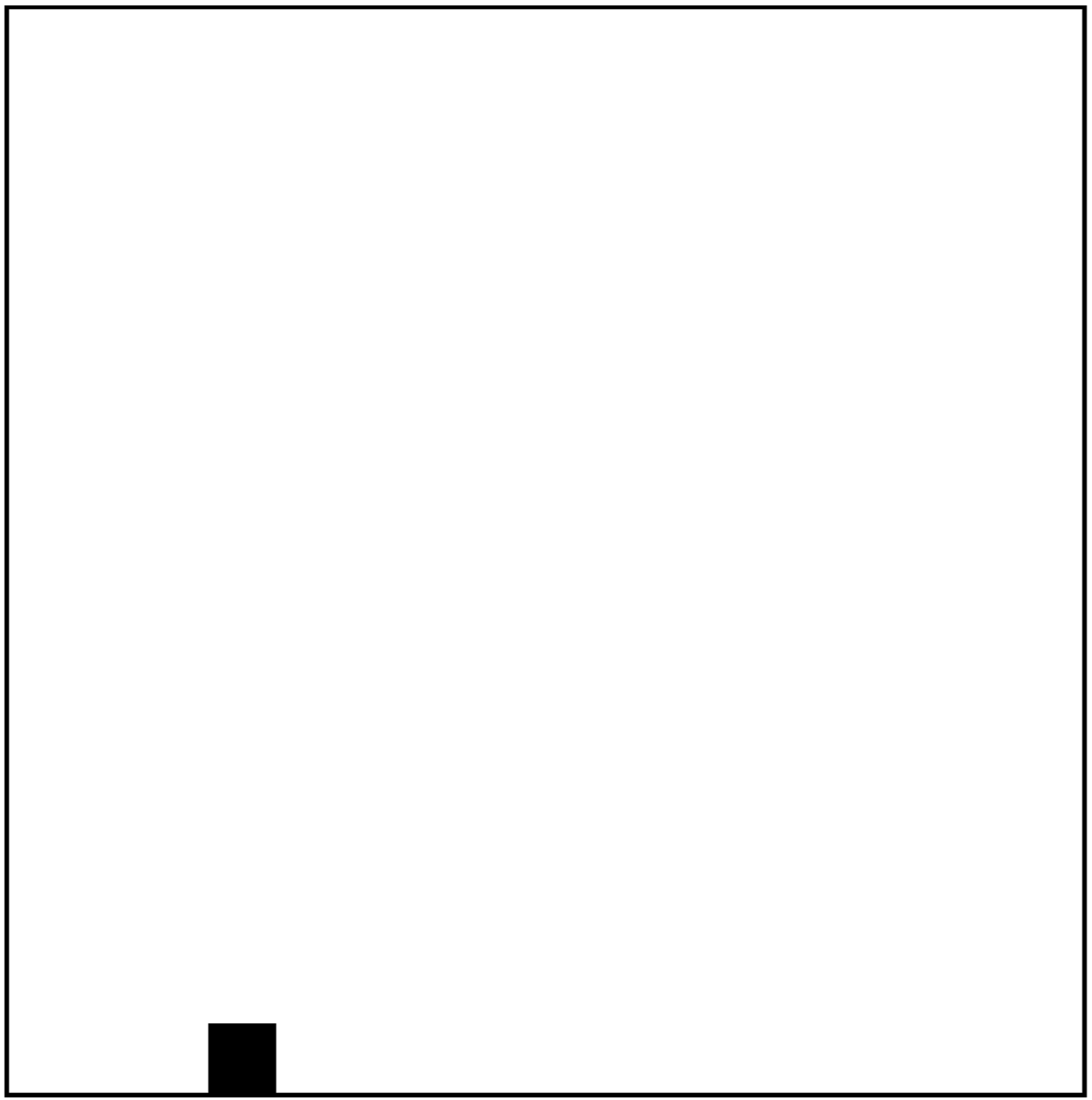} \\
a) & b) & c) & d) 
\end{tabular}
}
\caption{\label{fig:hshift} An example for hidden shift problem $g(x)=f(x\oplus s)$ over $A=\Z_2^{256}$ where the instance $f$ is given by a bent function and $f^\star$ is the dual bent function of $f$. Shown in a) is the circuit for the correlation-based algorithm from \cite{Roetteler:2010}, where the red marker denotes the state at the respective point in time during the algorithm's execution. Shown in b), c), and d) are visualizations of three stages during the algorithm's execution: b) is the state after the shifted function has been computed into the $\pm 1$-valued phase. Here black and light blue color stand for (re-normalized) values of $+1$ and $-1$, respectively. Shown in c) is the state after the Fourier transform. As bent functions have a flat spectrum in absolute value, the state is again two valued at this point. Finally, in d) the state after the final Hadamard transform is shown, after which all amplitude is supported on the shift $\ket{s}$. Here black denotes an amplitude of $1$ and white an amplitude of $0$. The main contribution of this paper, Algorithm \ref{alg:shiftedDiff}, can be considered a generalization of this picture to more general classes of hiding functions $f$. These are obtained from difference sets which, in a precise sense, generalize the notion of bent functions.}
\end{figure}

\subsection{Our results}

Based on the combinatorial structure of difference sets we derive a class of functions that have a two-level Fourier power spectrum. We then consider the hidden shift problem for these functions, following a general algorithm principle that was used earlier to solve the hidden Legendre symbol \cite{vDHI:2003} and the hidden bent function problem \cite{Roetteler:2010}. 

The basic idea underlying all these algorithms is to use the fact that the quantum computer can perform quantum Fourier transforms efficiently. This is used in correlation-based techniques which try to identify a shift by first transforming the function into frequency (Fourier) domain, then performing a point-wise multiplication with the desired target correlator, followed by an inverse Fourier transform and a measurement in the computational basis. After these steps the shift might be obtained, even without further post-processing. An example for this approach is shown in Figure \ref{fig:hshift} where the underlying group is the Boolean hypercube and the shifted function is a so-called bent function. 

A rich theory of difference sets exists and many explicit constructions are known. Furthermore, several applications of difference sets exist in signal processing \cite{PKHJ:98}, coding theory \cite{MS:77}, cryptography \cite{PHS:2003}, see also \cite{xBJL2:99} for further examples. In this paper we focus on the case of difference sets in abelian groups and show that a correlation-based approach can be successfully applied to several families of difference sets. In one application we consider so-called Singer difference sets, which are difference sets in cyclic groups. These difference sets have parameters 
\[
(v,k,\lambda) = \left( \frac{q^{d+1}-1}{q-1}, \frac{q^{d}-1}{q-1}, 
\frac{q^{d-1}-1}{q-1}\right),
\]
where $q$ is a constant and $d$ is a parameter that defines the input size of the problem, and $v$, $k$, and $\lambda$ are characteristic parameters of the difference set. We construct instances of dihedral hidden subgroup problems that can be (query) efficiently solved on a quantum computer. There is one step in our algorithm that requires the implementation of a diagonal operator whose diagonal elements are certain generalized Gauss sums whose flatness follows from a classic result due to Turyn \cite{Turyn:65}. In general, we do not known how to implement these diagonal operators efficiently and it seems that the actual computational cost has to be determined on a case-by-case basis. However, for the case of $q=2$, which leads to difference sets in a cyclic group of order $N=2^{d+1}-1$, we can leverage a result by van Dam and Seroussi \cite{vDS:2002} to implement a quantum algorithm that is fully efficient in terms of its quantum complexity as well as classical complexity. Classically, the underlying problem of this white-box problem is at least as hard as the 
discrete logarithm problem over a finite field. 

\subsection{Related work}

Several papers study the dihedral hidden subgroup, however, it is an open whether quantum computers can solve this problem efficiently. There is a quantum algorithm which is fully efficient in its quantum part, which however requires an exponential-time classical post-processing \cite{EH00}. Furthermore, a subexponential-time quantum algorithm for the dihedral subgroup problem is known~\cite{Kuperberg, Regev, Kuperberg2} based on a sieving idea. The dihedral hidden subgroup problem for adversarially chosen hiding functions is believed to be intractable on a quantum computer, even we are not aware of any evidence stronger for this intuition than reductions from lattice problems \cite{Regev:2004} and subset sum type problems \cite{Regev:2004,BCvD06}. The connection between hidden shift problems over abelian groups and hidden subgroup problems over semidirect groups of the mentioned special form is well-known and was one of the reasons why the hidden shift problem has been studied for various groups~\cite{EH00, vDHI:2003, Friedl2003, MRRS, CW07, Ivanyos:2008, CvD10}.

The study of hidden shift problems has resulted in quantum algorithms that are of independent interest and have even inspired cryptographic schemes that might be candidates for post-quantum cryptography \cite{Regev:2004b}. Besides the mentioned works, problems of hidden shift type were also studied in \cite{Roetteler:2009, GRR11, Gharibi:2013}, in the rejection sampling \cite{QRS} framework, and in the context of multiregister PGM algorithms for Boolean hidden shift problems \cite{CKOR:2013}. The main result of this paper is Theorem \ref{th:dihedral} which asserts that there exist instances of the hidden subgroup problem over the dihedral groups $D_N$ that can be solved in $O(\log N)$ queries to the hiding function, $O(polylog(N))$ quantum time, $O(\log N)$ quantum space, and trivial classical post-processing. Moreover, for $N=2^n-1$, where $n\geq 2$, there exist instances of the hidden subgroup problem over the dihedral group $D_{2^n-1}$ for which the hiding function is white-box and for which the entire quantum computation can be performed in $O(poly(n))$ quantum time, $O(n)$ quantum space, and trivial classical post-processing. Moreover, the classical complexity of solving these instances is at least as hard as solving the discrete logarithm problem over finite fields. To the best of our knowledge this is the first exponential size family of instances of the dihedral hidden subgroup problem that can be solved efficiently on a quantum computer, whereas for the same class of instances no efficient classical algorithm is known\footnote{It is easy to see that it is possible to find instances of the dihedral hidden subgroup problem that can be solved efficiently on a classical computer, e.g., functions that identify the points of a regular $N$-gon that are opposites along a symmetry axis in a linear increasing fashion. On these ``taco''-like instances the hidden symmetry axis, and thereby the hidden subgroup, can be found simply by a binary search.}. 

In Corollary \ref{cor:instances} we show that for $D_N$ where $N=2^n-1$ this theorem implies that there are an expected number of $O(2^{n^2})$ instances of the dihedral HSP (where the hidden subgroup is a reflection) that can be solved efficiently on a quantum computer. This is a small fraction of the set of all instances of such hidden subgroup problems as the number of all instances scales doubly exponential as $O(2^{n 2^n})$. In particular, it seems unlikely that the set of such constructed instances has a non-trivial intersection with the set of instances that can be obtain via Regev's reduction from gapped unique-SVP lattice problems. 

The rest of this paper is organized as follows. First, in Section \ref{sec:background} we introduce some notation and basic definitions such as Fourier transform, convolution, and the basic combinatorial object of study in this paper, namely difference sets in finite abelian groups. Next, in Section \ref{sec:qalgo} we present a quantum algorithm that can be applied to any shifted difference set problem, albeit sometimes with low probability of success. We exhibit some instances of shifted difference set problems that can be solved efficiently. These special cases include the so-called class of Singer difference sets which are then used in Section \ref{sec:dihedral} to construct instances of the dihedral hidden subgroup problem that can be solved efficiently on a quantum computer. Finally, in Section \ref{conclusions} we offer conclusions and end with some open problems. 

%
%

\section{Background}\label{sec:background}

\subsection{Quantum Fourier transforms over abelian groups} \label{sect:Notation}

The main tool we will use are Fourier transforms over abelian groups. In the following we state some basic definitions and properties. Recall that for any abelian group $A$ the character group $\widehat{A} = {\rm Hom}(A,\C^\times)$ is isomorphic to $A$. We denote the irreducible characters of $A$ by $\chi: A \rightarrow \C^\times$. 

\begin{definition}
The \emph{quantum Fourier transform} on $\C^d$ is a unitary transformation defined as $\QFT_A := \frac{1}{\sqrt{|A|}} \sum_{a\in A} \sum_{\chi\in\widehat{A}} \chi(a) \ket{\chi}\bra{a}$.
\end{definition}

\begin{example} For $A=\Z_2$ the $\QFT_A$ is given by the Hadamard transform $H := \frac{1}{\sqrt{2}} \bigl( \begin{smallmatrix*}[r] 1 & 1 \\ 1 & -1 \end{smallmatrix*} \bigr)$. 
\end{example}

\begin{definition}\label{def:Fourier}
The \emph{Fourier transform} of a (complex-valued) function $F: A \to \C$ is a function $\hat{F}: \widehat{A} \to \C$ defined as $\hat{F}(\chi) := \bra{\chi} \QFT_A \ket{F}$ where $\ket{F} := \sum_{x \in A} F(x) \ket{x}$. Here $\hat{F}(\chi)$ is called the \emph{Fourier coefficient} of $F$ at $\chi\in\widehat{A}$. We can write it explicitly as $\hat{F}(\chi) = \frac{1}{\sqrt{|A|}} \sum_{x \in A} \chi(x) F(x)$. The set $\{\hat{F}(\chi) : \chi \in \widehat{A}\}$ is called the \emph{Fourier spectrum} of $F$.
\end{definition}

\begin{definition}
The \emph{convolution} of functions $F, G: A \to \C$ is a function $(F * G): A \to \C$ defined as $(F * G)(x) = \sum_{y \in A} F(y) G(x - y)$.
\end{definition}

\begin{fact}\label{fact:Basics}
Let $F,G,H: A \to \C$ denote arbitrary functions. The Fourier transform and convolution have the following basic properties:
\begin{enumerate}
  \item The Fourier transform is linear: $\widehat{F+G} = \hat{F} + \hat{G}$.
  \item When applied twice, the Fourier transform satisfies $\hat{\hat{F}}(z) = F(-z)$. In particular, for $A=\Z_2$ the Fourier transform is self-inverse: $\hat{\hat{F}} = F$. From this property also follows that when the Fourier transform $\QFT_A$ is applied four times then the result is the identity. 
  \item $\QFT$ is unitary, so the Plancherel identity $\sum_{\chi \in A} |\hat{F}(\chi)|^2 = \sum_{x \in A} |F(x)|^2$ holds.
  \item The convolution is commutative: $F * G = G * F$, and associative: $(F * G) * H = F * (G * H)$.
  \item The Fourier transform and convolution are related through the following identities: $(\hat{F} * \hat{G}) / \sqrt{|A|} = \widehat{FG}$ and $(\widehat{F * G}) / \sqrt{|A|} = \hat{F} \hat{G}$, where $FG: A \to \C$ is the entry-wise product of functions $F$ and $G$: $(FG)(x) := F(x) G(x)$.
\item A shift of a function in time domain leads to a point-wise multiplication with a ``linear phase'' in Fourier domain: If there exists $s\in A$ such that for all $x\in A$ it holds $G(x) = F(x-s)$, then for all $\chi \in \widehat{A}$ we have that $\hat{G}(\chi) = \chi(s) \hat{F}(\chi)$. This latter property will be crucial for the hidden shift algorithm presented later in this paper. 
\end{enumerate}
\end{fact}

\subsection{Difference sets}

We recall the definition of difference sets in finite groups. We focus on the case of abelian groups in this paper. See also \cite{BJL1:99,Stinson:2003,Ladner:83} for further information, in particular about the treatment for general, non-abelian groups. 

Let $A$ be a finite abelian group whose group operation we write additively and whose neutral element we denote with $0_A$. Denote the pairwise inequivalent irreducible characters of $A$ by  $\widehat{A}$. For a subset $D \subseteq A$ 
of $A$ we introduce the notation $D^-  := \{ -d : d \in D\}$ for the set of all inverses and $\Delta D := D+D^- = \{ x - y : x, y \in D\}$ for the set of all differences of pairs of elements of $D$.  

\begin{definition}[Difference set]
Let $A$ be a finite abelian group of size $v = |A|$. A subset $D \subseteq A$ of size $k = |D|$ is called a $(v, k, \lambda)$-difference set, where $\lambda \geq 1$, if the following equality holds in the group algebra $\C[A]$ of $A$:
\begin{equation}\label{eq:diffSet}
\Delta D = \lambda (A \setminus \{0_A\}) + k 0_A.
\end{equation}
\end{definition}
This means that the set of all differences covers each element the same number $\lambda$ of times, except for the neutral element, which is covered precisely $k$ times. A nice feature of difference sets in abelian group is that they allow to construction functions with almost flat spectrum: the following theorem \cite{Turyn:65} asserts that all Fourier coefficients of the characteristic function of a difference set in an abelian group have the same absolute value, with a possible exception of a peak at the zero frequency: 

\begin{theorem}[Turyn, 1965]\label{th:turyn}
Let $A$ be an abelian group of order $v$ and $D$ be an $(v, k, \lambda)$-difference set in $A$. Let $\chi \in \widehat{A}$ be a non-trivial character. Then 
\begin{equation}\label{eq:charsum}
|\chi(D)| := \left| \sum_{d \in D} \chi(d) \right| = \sqrt{k-\lambda}
\end{equation}
holds. For the trivial character $\chi_0$ we have that $|\chi_0(D)| = k$.
\end{theorem}

\proof
We include a proof as it is instructive to see how the difference set condition  can be used when interpreted as the identity (\ref{eq:diffSet}) in the group ring $\C[A]$. Indeed, when identifying $D$ with $\sum_{d \in D} d \in \C[A]$, we obtain from eq.~(\ref{eq:diffSet}) that 
\begin{equation}\label{eq:char}
\left(\sum_{d\in D} d\right) \left(\sum_{d\in D} -d\right) = \lambda\left(\sum_{g \in A} g \right) + (k-\lambda) 0_A. 
\end{equation}
Let $\chi\in\widehat{A}$ be non-trivial. Then clearly $\chi(A)=0$ holds which implies---by applying $\chi$ to both sides of eq.~(\ref{eq:char})---that $\chi(D) \overline{\chi(D)} = \chi(A) + (k-\lambda) \chi(0_A) = k-\lambda$. From this we obtain that $|\chi(D)| = \sqrt{k-\lambda}$ as claimed. \qed

With each difference set $D$ we can canonically associate an incidence structure called the {\em development} of $D$, and denoted by $Dev(D)$.  

\begin{definition} Let $D$ be a $(v,k,\lambda)$-difference set in an abelian group $A$. Then the points of $Dev(D)$ are given by the elements of $A$ and the blocks of $Dev(D)$ are given by $v+D := \{ v + a : a \in D\}$, where $v\in A$. 
\end{definition} 

It is well-known that $Dev(D)$ is a symmetric design. More precisely, we have the following result (for a proof see, e.g., \cite{BJL1:99}, \cite{Stinson:2003} or \cite{Ladner:83}): 

\begin{theorem}\label{th:symmetric}
Let $D$ be a $(v,k,\lambda)$-difference set in an abelian group $A$. Then $Dev(D)$ is a symmetric balanced-incomplete block design with parameters $(v,k,\lambda)$. 
\end{theorem}

Theorem \ref{th:symmetric} implies that there are $|A|$ blocks, that each block has $|D|$ elements, that any two elements have precisely $\lambda$ blocks in common and that in addition any two blocks intersect in precisely $\lambda$ points. Also, it holds that $\lambda=k(k-1)/(v-1)$, see e.g. \cite[Prop.~1.1]{Ladner:83}, implying that $\lambda$ is determined by the group order $v$ of $A$ and the size $k$ of $D$. This equality allows us also to do a consistency check that the normalized state vector $\frac{1}{\sqrt{k}} \sum_{d \in D} \ket{d}$ is indeed mapped to a normalized vector under the Fourier transform ${\rm QFT}_A$ for the group $A$: using Theorem \ref{th:turyn} we find that the length of the transformed vector is given by 
\begin{eqnarray*}
\frac{1}{\sqrt{vk}^2} \left( (v-1) |\chi(D)|^2 + |\chi_0(D)|^2 \right) 
&=& ((v-1)(k-\lambda) + k^2)/(vk) \\
&=& ((v-1)k - k (k-1) + k^2)/(vk) =1,
\end{eqnarray*}
as desired.  

%
%

\section{Quantum algorithm for shifted difference sets}
\label{sec:qalgo}

\begin{problem}[Shifted difference set problem]\label{def:membership}
Let $A$ be an abelian group and let $s \in A$. Let $D \subseteq A$ be a (known) difference set and let $s+D$ be given by a membership oracle. The problem is to find $s$. 
\end{problem}

\begin{figure}[hbt]
  \begin{center}
 \includegraphics[width=6cm]{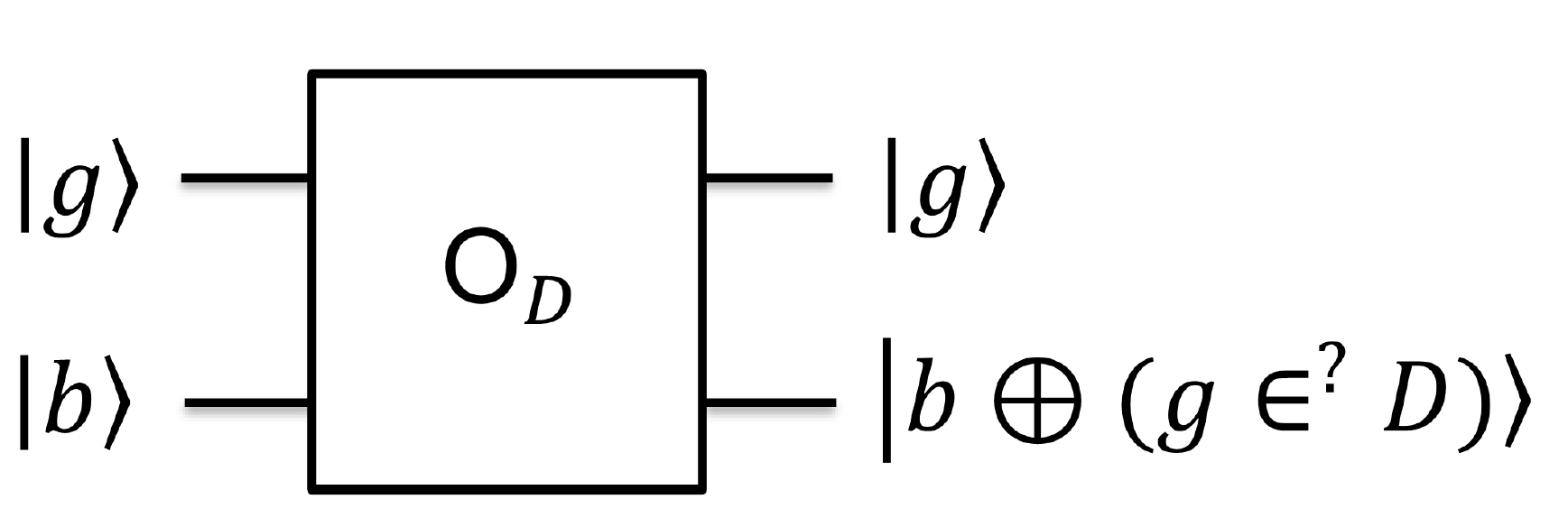}
  \caption{  \label{fig:oracle} The oracle for the shifted difference set problem considered in this paper. The oracle allows to test membership of a given test input $g \in A$. The value of the test $g \in^? D$ is XORed onto a bit $b \in \{0,1\}$.}
\end{center}
\end{figure}

Similar to \cite{Roetteler:2010} we can modify Problem \ref{def:membership} by hiding not only the characteristic function of $s+D$ via a membership oracle but also the characteristic function of $D$ itself. In this case we assume that we have access to membership oracles for both $D$ and $s+D$. The following quantum algorithm is a general recipe to tackle instances of the shifted difference set problem specified in Problem \ref{def:membership}. As we will show in the following, Algorithm \ref{alg:shiftedDiff} can be used to find the hidden shift $s$ efficiently in several cases of difference sets for various abelian groups $A$. It should be noted, however, that the probability of success crucially depends on the instance $(A, D)$ of the problem and there are instances for which the algorithm recovers $s$ successfully is only exponentially small. The algorithm can be seen as a generalization of correlation-based algorithms for solving hidden shift problems, e.g., \cite{vDHI:2003}, \cite{Roetteler:2010}, and \cite{CKOR:2013}.

\begin{algorithm}\label{alg:shiftedDiff}
The input to the algorithm is a membership oracle as in Problem  \ref{def:membership}. 

\noindent {\bf Step 1:} Prepare the input superposition: 
\[
\ket{0} \mapsto \frac{1}{\sqrt{|A|}} \sum_{g \in A} \ket{g}.
\]
\noindent {\bf Step 2:} Query the shifted difference set. This maps the state to:
\[
\frac{1}{\sqrt{|A|}} \sum_{g \in A} (-1)^{(g \in^? s+D)} \ket{g}
= \frac{1}{\sqrt{|A|}} \sum_{g \in A} \ket{g} - \frac{2}{\sqrt{|A|}} \sum_{d \in (s+D)} \ket{d}.
\]
\noindent {\bf Step 3:} Apply the quantum Fourier transform for $A$. This maps the state to:
\[
\ket{\chi_0} - \frac{2}{|A|} \sum_{\chi \in \widehat{A}} \chi(s+D) \ket{\chi}
= \left( 1-\frac{2k}{|A|}\right)\ket{\chi_0} - \frac{2}{|A|} \sum_{\chi \not= \chi_0} \chi(D) \chi(s)\ket{\chi}.
\]
\noindent {\bf Step 4:} Compute ${\rm diag}(1,\overline{\chi(D)}/\sqrt{k-\lambda} : \chi\not=\chi_0)$ into the phase. This maps the state to: 
\[
\left( 1-\frac{2k}{|A|}\right)\ket{\chi_0} - \frac{2(k-\lambda)}{|A|} \sum_{\chi \not= \chi_0} \chi(s) \ket{\chi}.
\]
\noindent {\bf Step 5:} Apply the inverse quantum Fourier transform for $A$. This maps the state to: 
\[
\frac{1}{\sqrt{|A|}} \left( 1-\frac{2(k-\sqrt{k-\lambda})}{|A|}\right) \sum_{g \in A} \ket{g} - \frac{2 \sqrt{k-\lambda}}{\sqrt{|A|}} \ket{-s}
\]
\noindent {\bf Step 5:} Measure in the standard basis. 
Obtain $-s$ with probability $p := \frac{4 (k-\lambda)}{|A|}$ and all other group elements uniformly with probability $(1-p)/|A|$. 
\end{algorithm}

\subsection{Examples}

\subsubsection{Paley difference sets and shifted Legendre functions}
Let $A$ be the additive group of the finite field $\F_q$, where $q=p^n$ is a prime power such that $q \equiv 3 \; ({\rm mod} \, 4)$. Define $D := \{ x : x \; \mbox{is a non-zero square in} \; \F_q \}$. It is well-known \cite{BJL1:99,Stinson:2003} that $D$ is then a difference set in $A$. These difference sets are also known as Paley difference sets. The parameters of $D$ are as follows: 
\[
(v,k,\lambda) = \left( q, \frac{q-1}{2}, \frac{q-3}{4} \right).
\]

\begin{example} Let $q=27$ and consider the irreducible polynomial $f(x) = x^3+x^2+x+2 \in \F_3[x]$, defining the finite field $\F_{27} \cong \F_3[x]/(f(x))$. Denote by $\{1, \alpha, \alpha^2\}$ an $\F_3$-basis of $\F_{27}$ where $\alpha := x \; {\rm mod}\; f(x)$, the image of $x$ under the canonical projection. Then $D$ given by the following $13$ elements
\begin{eqnarray*}
D &=& \{ 
1, \quad \alpha, \quad 2\alpha^2 + 2\alpha + 1, \quad 2\alpha + 2, \quad \alpha + 2, \quad \alpha^2 + 2\alpha, \quad \alpha^2 + 1, \quad 2\alpha^2 + 1,\\
&& \;\;\alpha^2 + \alpha + 1, \quad \alpha^2, \quad 2\alpha^2 + 2\alpha, \quad \alpha^2 + 2\alpha + 1, \quad \alpha^2 + 2\alpha + 2 \}
\end{eqnarray*}
defines a $(27,13,6)$-difference set in $\Z_3^3$. 
\end{example}
\begin{remark}
Applying Algorithm \ref{alg:shiftedDiff} finds the hidden shift with probability of success 
\[
p_{success} = \left|  \frac{2 \left( \frac{q-1}{2} - \frac{q-3}{4} \right)^{1/2}}{q^{1/2}} \right|^2 \approx 1-O(1/q).
\]
This means that for large $q$, we can efficiently recover the hidden shift. In this case, the Algorithm \ref{alg:shiftedDiff} specializes to the algorithm given in \cite{vDHI:2003}. We recover the result that an unknown shift of the Legendre symbol can be reconstructed with high probability using $1$ query. 
\end{remark}

\subsubsection{Hadamard difference sets and shifted bent functions}
Let $A$ be the elementary abelian $2$-group $A=\Z_2^{2n}$, where $n\in\N$. Let $f: \Z_2^{2n} \rightarrow \Z_2$ be a bent function. Define $D := \{ x\in \Z_2^{2n} :f(x) = 1 \}$. It is well-known \cite{BJL1:99,Stinson:2003} that $D$ is a difference set in $A$. These difference sets are also known as Hadamard difference sets. The parameters of $D$ are as follows: 
\[
(v,k,\lambda) = \left(2^{2n},\; 2^{2n-1}-2^{n-1}, \;2^{2n-2}-2^{n-1}\right).
\]

\begin{example} Let $n=4$ and let $f(x_1,x_2,x_3,x_4) = x_1 x_2 \oplus x_3 x_4 \oplus x_1 \in \F_2[x_1, x_2, x_3, x_4]$ be a bent function from the Maiorana-McFarland family \cite{Dillon72}. Then $D = \{ x \in \F_2^4 : f(x) = 1 \}$  given by the following 
$6$ elements
\begin{eqnarray*}
D &=& \{ (1, 1, 0, 0), \quad (1, 1, 1, 0), \quad (1, 1, 0, 1), \quad (0, 0, 1, 1), \quad (1, 0, 1, 1), \quad (0, 1, 1, 1) \}
\end{eqnarray*}
defines a $(16,6,2)$-difference set in $\Z_2^4$. The blocks of the development $Dev(D)$ of $D$ are obtained by taking the characteristic function of $f$ and shifting it under all elements of $A = \Z_2^4$. Hence, the incidence matrix of the $(16,6,2)$-design $Dev(D)$ is given by 
\[
\left[
\begin{array}{cccccccccccccccc}
     0& 0& 0& 1& 0& 0& 0& 1& 0& 0& 0& 1& 1& 1& 1& 0 \\
     0& 0& 1& 0& 0& 0& 1& 0& 0& 0& 1& 0& 1& 1& 0& 1 \\
     0& 1& 0& 0& 0& 1& 0& 0& 0& 1& 0& 0& 1& 0& 1& 1 \\
     1& 0& 0& 0& 1& 0& 0& 0& 1& 0& 0& 0& 0& 1& 1& 1 \\
     0& 0& 0& 1& 0& 0& 0& 1& 1& 1& 1& 0& 0& 0& 0& 1 \\
     0& 0& 1& 0& 0& 0& 1& 0& 1& 1& 0& 1& 0& 0& 1& 0 \\
     0& 1& 0& 0& 0& 1& 0& 0& 1& 0& 1& 1& 0& 1& 0& 0 \\
     1& 0& 0& 0& 1& 0& 0& 0& 0& 1& 1& 1& 1& 0& 0& 0 \\
     0& 0& 0& 1& 1& 1& 1& 0& 0& 0& 0& 1& 0& 0& 0& 1 \\
     0& 0& 1& 0& 1& 1& 0& 1& 0& 0& 1& 0& 0& 0& 1& 0 \\
     0& 1& 0& 0& 1& 0& 1& 1& 0& 1& 0& 0& 0& 1& 0& 0 \\
     1& 0& 0& 0& 0& 1& 1& 1& 1& 0& 0& 0& 1& 0& 0& 0 \\
     1& 1& 1& 0& 0& 0& 0& 1& 0& 0& 0& 1& 0& 0& 0& 1 \\
     1& 1& 0& 1& 0& 0& 1& 0& 0& 0& 1& 0& 0& 0& 1& 0 \\
     1& 0& 1& 1& 0& 1& 0& 0& 0& 1& 0& 0& 0& 1& 0& 0 \\
     0& 1& 1& 1& 1& 0& 0& 0& 1& 0& 0& 0& 1& 0& 0& 0 
\end{array}
\right].
\]
\end{example}
Applying Algorithm \ref{alg:shiftedDiff} to the shifted difference problem for a Hadamard difference set finds the hidden shift with probability of success 
\[
p_{success} = \left| \frac{2 \left( 2^{2n-1}-2^{2n-2}\right)^{1/2}}{{2^{2n}}^{1/2}} \right|^2 = 1.
\]
This means that we always recover the hidden shift $s$ with probability $1$. In this case, the Algorithm \ref{alg:shiftedDiff} specializes to the algorithm given in \cite{Roetteler:2010}. We recover the result that an unknown shift of a bent function can be reconstructed using $1$ query. 

\subsubsection{Singer difference sets and shifted hyperplanes}
Let $q$ be a prime power, let $d\geq 1$ and let $\F_{q^{d+1}}$ be the finite field with $q^{d+1}$ elements. The Singer difference sets are constructed from $d$-dimensional projective spaces over $\F_q$ as follows: consider the trace map ${\rm tr}$ from $\F_{q^{d+1}}$ to $\F_q$. Let $T$ be a transversal of $\F_q^*$ in $\F_{q^{d+1}}^*$ that is chosen in such a way that ${\rm tr}$ maps $T$ onto the values $0$ and $1$ in $\F_q$ only. We can then define a group $A := \F_{q^{d+1}}^\times/\F_q^\times$ which turns out to be cyclic. Furthermore, we can define a subset $D := \{x : x \in A | {\rm tr}(x) = 0\}$. It turns out \cite{BJL1:99} that $D$ is then a difference set in $\Z_N$, where $N = \frac{q^{d+1}-1}{q-1}$. This difference set has parameters 
\begin{equation}\label{eq:singerpars}
(v,k,\lambda) = \left( \frac{q^{d+1}-1}{q-1}, \frac{q^{d}-1}{q-1}, 
\frac{q^{d-1}-1}{q-1}\right). 
\end{equation}

\begin{example} 
Let $P = PG(2,3)$ be the two-dimensional projective space over $\F_3$. Then $|P| = (3^3-1)/(3-1) = 13$. By choosing an $\F_3$-basis of $\F_{27}$ we obtain an embedding of $\F_{27}^\times$ into ${\rm GL}(3,\F_3)$. If $\alpha\in \F_{27}$ is a primitive element for $\F_{27}/\F_3$, then the corresponding matrix has order $26$ and therefore generates a cyclic subgroup $C$ of ${\rm GL}(3,\F_3)$ order $26$. Under the canonical projection $\pi: {\rm GL}(3,\F_3) \rightarrow {\rm PGL}(3,\F_3)$, the subgroup $C$ is mapped to a subgroup $\overline{C}=\langle \sigma \rangle$ of ${\rm PGL}(3,\F_3)$ of order $13$ (see also \cite[Kapitel II, Satz~7.3]{Huppert:67}). This subgroup is sometimes also called the ``Singler cycle.'' The Singer cycle operates transitively on the points $\{(x:y:z) : x,y,z \in \F_3 \}$ of the projective space $P$. By picking the particular order $[\sigma^i p_0 : i=0, \ldots, 12]$, where $p_0$ is the point $(0:0:1)$, we obtain points that we can identify with $[0,1,\ldots,12]$. The image of the hyperplane given by all points $p \in \F_{27}$ with ${\rm tr}(p) = 0$ is given by the set $D := \{0,1,3,9\}$. Then $D$ is a $(13,4,1)$-difference set in the cyclic group $\Z_{13}$. The development $Dev(D)$ is given by:
\[
\left[
\begin{array}{ccccccccccccc}
1& 0& 0& 0& 1& 0& 0& 0& 0& 0& 1& 0& 1\\
1& 1& 0& 0& 0& 1& 0& 0& 0& 0& 0& 1& 0\\
0& 1& 1& 0& 0& 0& 1& 0& 0& 0& 0& 0& 1\\
1& 0& 1& 1& 0& 0& 0& 1& 0& 0& 0& 0& 0\\
0& 1& 0& 1& 1& 0& 0& 0& 1& 0& 0& 0& 0\\
0& 0& 1& 0& 1& 1& 0& 0& 0& 1& 0& 0& 0\\
0& 0& 0& 1& 0& 1& 1& 0& 0& 0& 1& 0& 0\\
0& 0& 0& 0& 1& 0& 1& 1& 0& 0& 0& 1& 0\\
0& 0& 0& 0& 0& 1& 0& 1& 1& 0& 0& 0& 1\\
1& 0& 0& 0& 0& 0& 1& 0& 1& 1& 0& 0& 0\\
0& 1& 0& 0& 0& 0& 0& 1& 0& 1& 1& 0& 0\\
0& 0& 1& 0& 0& 0& 0& 0& 1& 0& 1& 1& 0\\
0& 0& 0& 1& 0& 0& 0& 0& 0& 1& 0& 1& 1\\
\end{array}
\right].
\]
\end{example}
If in eq.~(\ref{eq:singerpars}) we consider $q$ to be constant and $d$ be a parameter that corresponds to the input size of a hidden shift problem over $\Z_N$, we can use Algorithm \ref{alg:shiftedDiff} to solve the hidden shift problem over $\Z_N$ with probability of success 
\[
p_{success} = \left|  \frac{4 (q^d-q^{d-1})^{1/2}}{(q^{d+1}-1)^{1/2}} \right|^2 = \frac{2}{q} + O(1/q^2).
\]
This means that for constant $q$, we can efficiently recover the hidden shift from a constant number of trials. In Section \ref{sec:dihedral} we show how we can use the instances of hidden difference problems of Singer type to construct efficiently solvable instances of the dihedral hidden subgroup problem. 

\begin{remark} We note that not all shifted difference set problems can be solved efficiently by using Algorithm \ref{alg:shiftedDiff}. An example is given by the projective planes $(q^2+q+1,q+1,1)$ of order $q$. In this case the input size is given by $\log{q}$ and the probability of success can be computed to be $p_{success} = |\frac{2 q^{1/2}}{(q^2+q+1)^{1/2}}|^2 \approx \frac{2}{q} + O(1/q^2)$, i.e., the probability of success is exponentially small in this case. It is an open problem if cases like this can be tackled, e.g., by considering multi-register algorithms. 
\end{remark}

\subsection{Injectivization}\label{sec:inj}

As mentioned in the introduction, it is well known that the hidden subgroup problem over semidirect products of the form $A \rtimes \Z_2$, where the action of $\Z_2$ is given by inversion, and the hidden shift problem over $A$ are closely related. More precisely, there is a one-to-one correspondence between instances of the hidden subgroup problem in which the subgroup is a conjugate of the order $2$ subgroup $H = \langle (0,1) \rangle$ and instances of {\em injective} hidden shift problems over $A$. 

This leads to the question whether it is possible to relate instances of hidden shift problems where the hiding function $f : A \rightarrow S$ is not injective to the injective case. Thankfully, as shown in \cite{Gharibi:2013} such a connection indeed exists. We briefly review this construction. 

For given $f : A \rightarrow S$, and a set $V := \{v_1, \ldots, v_m\} \subseteq A$ of $m$ elements of $A$ we define a new function $f_V(x) := (f(x+v_1), \ldots, f(x+v_m))$. Gharibi showed in \cite{Gharibi:2013} that if the set $V$ is chosen uniformly at random, then the probability that the function $f_V$ is not injective can be upper bounded as 
\begin{equation}\label{eq:injbound}
Pr_V(f_V \; \mbox{not injective}) \leq |A|^2(1-\gamma_{min})^m,
\end{equation}
where $\gamma_{min} := \min_{v\not= 0}(\gamma_v(f))$ and for all $v \in A$ the so-called influences $\gamma_v(f)$ of $f$ at $v$ are defined as $\gamma_v(f) := Pr_x (f(x) \not= f(x+v))$, i.e., the probability that $f$ changes its value when the input is toggled by $v$. 

We now show that for instances of shifted difference set problems these influences can be bounded by the parameters of the difference set alone. This in turn allows to establish a bound on the overall number of copies $m$ that are needed to make the hiding function injective, namely a bound that grows proportional to $\log |A|$. 

\begin{lemma}\label{lem:bound} Let $f: A \rightarrow \{0,1\}$ be a hiding function corresponding to the characteristic function a $(v,k,\lambda)$-difference set in an abelian group $A$. Then for all $v \in A\setminus\{0\}$ we have that $\gamma_v(f) = \frac{2(k-\lambda)}{|A|}$. 
\end{lemma}
\proof 
Note that 
\begin{eqnarray}
Pr_x(f(x) \not= f(x+v)) &=& \frac{1}{|A|} \sum_{x \in A}\left(f(x)-f(x+v)\right)^2\\
&=& \frac{1}{|A|} \left( |D| + |v+D| - 2 |D \cap (v+D)| \right) \\
&=& \frac{1}{|A|} (2|D|-2\lambda) = \frac{2(k-\lambda)}{|A|},
\end{eqnarray}
where in the second equation we used the fact that only elements in the intersection contribute to $f(x)f(x+v)$ and in the third equation we used Theorem \ref{th:symmetric} which implies that $|D \cap (v+D)|=\lambda$ for all $v\not=0$. 
\qed

We can now establish the claimed result that the number of copies only grows with the log of the group size. 

\begin{theorem}\label{th:injectivize}
Let $D$ be a $(v,k,\lambda)$-difference set in an abelian group $A$ and $f: A \rightarrow \{0,1\}$ an instance of a hidden difference set problem for $D$. Then $m=O(\log{|A|})$ copies are enough to obtain an injective instance $f_V$ with probability greater than $1-\frac{1}{64}$. 
\end{theorem}
\proof
From the cited bound (\ref{eq:injbound}) we obtain that 
\[
Pr(f \; \mbox{injective}) \geq  1-{|A|}^2(1-\gamma_{min}(f))^m
\]
It is easy to see that lower bounding the right hand side in this expression by $1-\frac{1}{64}$ is equivalent to choosing $m \geq \frac{1}{\log(1-\gamma_{min}(f))}(-6-2 \log_2(|A|))$. Now, from Lemma \ref{lem:bound} we have that $\gamma_{min}(f) = \frac{2(k-\lambda)}{|A|}$ from which we can conclude that in particular $|A|\geq 2(k-\lambda)$ holds. Using the fact that $\log_2(1-x)\leq -x$ holds for $x\in [0,1)$, this implies that 
\begin{eqnarray}
m &\geq& \Big\lceil\frac{1}{\log_2\left(1-\frac{2(k-\lambda)}{|A|}\right)}(-6-2\log_2{|A|})\Big\rceil \\
&\geq& \Big\lceil - \frac{|A|}{2(k-\lambda)}(-6 -2 \log_2{|A|})\Big\rceil \geq 
2 \log_2{|A|}+6.
\end{eqnarray}
Hence $m=O(\log{|A|})$ copies are enough to guarantee that for $V=\{v_1, \ldots v_m\}$ chosen uniformly at random, the probability of $f_V$ being injective is at least $1-\frac{1}{64}$. 
\qed

%
%

\section{Efficiently solvable dihedral hidden subgroup problems}
\label{sec:dihedral}

\begin{theorem}\label{th:dihedral} There exist instances of the hidden subgroup problem over the dihedral groups $D_N$ that can be solved in $O(\log N)$ queries to the hiding function, $O(polylog(N))$ quantum time, $O(\log N)$ quantum space, and trivial classical post-processing. Moreover, for $N=2^n-1$, where $n\geq 2$, there exist instances of the hidden subgroup problem over the dihedral group $D_{2^n-1}$ for which the hiding function is white-box and for which the entire quantum computation can be performed in $O(poly(n))$ quantum time, $O(n)$ quantum space, and trivial classical post-processing. Moreover, the classical complexity of solving these instances is at least as hard as solving the discrete logarithm problem over finite fields. 
\end{theorem}

\proof
To construct the instances that can be solved efficiently we proceed in three steps: (i) first, we show that a particular set of hidden shift problems over $\Z_N$ can be obtained from hiding functions that are indicator functions of hyperplanes and that these indicator functions can be implemented efficiently, (ii) next we show that Algorithm \ref{alg:shiftedDiff} is query, time, and space efficient for these instances; (iii) finally, we show that it is possible to construct instances of the hidden subgroup problem in $D_N$ from the hidden shift instances constructed in (i) and that these instances are unlikely to be solvable on a classical computer, unless computing finite field discrete logarithms is possible in polynomial-time. 

Step (i): We instantiate the abelian difference set quantum algorithm for the case of the cyclic group $A=\Z_N$, where $N=(q^{d+1}-1)/(q-1)=q^d+q^{q-1}+\ldots+1$. Here $q$ is constant and $d$ is a parameter that corresponds to the input size of the problem. We use the explicitly (white-box) description of the function $f(x) = {\rm tr}(\alpha^x)$, where ${\rm tr}$ denotes the trace map from $\F_q^{d+1}$ to $\F_q$ and where $\alpha$ is a primitive element in $\F_q$. Now, the instance of the shifted difference set problem is defined by the hiding function  $g(x) = {\rm tr}(\alpha^{x+s})$, where $s \in \Z_N$. This function can be given as a white-box function by providing the element $\beta := \alpha^s \in \F_{q^{d+1}}$ so that $g$ can then be evaluated as $g(x) = {\rm tr}(\alpha^x \beta)$. Note that the set $\{x \in A : {\rm tr}(x) = 0\}$ defines a hyperplane and therefore a difference set $D$ of Singer type. 

Step (ii): We now go through each step of Algorithm \ref{alg:shiftedDiff} and check that the steps are time- and space-efficient. In the first step, a Fourier transform is applied to create the equal superposition of all elements of $A$. As $A$ is abelian, this can clearly be done efficiently. In the second step, we have to evaluate the function $g$ in superposition. Again, as there is an explicit description of the trace which can be computed as sum of powers of the relative Frobenius from $\F_{q^{d+1}}$ to $\F_q$ as follows ${\rm tr}(x) = x + x^q + \ldots + x^{q^{d}}$ we can evaluate $g(x) = {\rm tr}(\alpha^x\beta)$ by first constructing a circuit for exponentiation $x \mapsto \alpha^x \in \F_{q^{d+1}}$ followed by scalar multiplication with $\beta$, followed by the application of the trace map. Clearly, all these operations can be efficiently implemented by means of a classical Boolean circuit whose size and depth are polynomial in $d$. Hence, by applying standard techniques from reversible computing, we can derive quantum circuits for the evaluation of $f$ and $g$. Therefore we can compute Step 2 efficiently on a quantum computer. 

Step 3 is another application of a quantum Fourier transform over the abelian group $A$ which as in Step 1 can be done efficiently. Step 4 is the most challenging step in the entire algorithm. If we were just interested in the query complexity of the problem we would be done as we could simply apply the diagonal unitary operator $\Delta := {\rm diag}(1,\overline{\chi_1(D)}, \ldots, \overline{\chi_{N-1}(D)})$, where $\chi_1, \ldots, \chi_{N-1}$ runs through all non-trivial characters of $\Z_N$. This argument is sufficient to establish the first claimed statement in the theorem, i.e., the query complexity result. 

For the white-box statement, we are interested in the time- and space-efficiency of the algorithm, i.e., we have to show that $\Delta$ can be implemented efficiently. For this we have to assume $N=2^n-1$ as required by one of the subsequent steps (and we highlight where). First we use a result due to van Dam and Seroussi \cite{vDS:2002} establishing that finite field Gauss sums can be approximated efficiently on a quantum computer. The connection to our situation is that the elements of $\Delta$ are Gauss sums. We briefly review the van Dam/Seroussi algorithm and then argue that we can apply it in superposition in order to compute $\Delta$. 

Let $\F_q$ be a finite field where $q=p^{d+1}$ and $p$ prime. Let $\psi := \F_p \rightarrow \C^\times$ be a non-trivial additive character and let $\chi: \F_q^\times \rightarrow C^\times$ be a non-trivial multiplicative character. Then the Gauss sum $G(\psi,\chi)$ is defined as 
\[
G(\psi,\chi) = \sum_{x \in \F_q^\times} \chi(x) \psi({\rm tr}(x)).
\]
The additive and multiplicative characters of $\F_q$ have a simple description: For $n\in \N$ denote a primitive $n$-th root of unity in $\C^\times$ with $\omega_n$. Then the additive characters take the form $\psi_\mu(x) := \omega_p^{{\rm tr}(\mu x)}$, where $\mu \in \F_q$ runs through all elements of $\F_q$. The multiplicative characters can be described using a primitive elements $\alpha \in \F_{p^{d+1}}$ as follows: $\chi_\beta(\alpha^i) := \omega_{p^{d+1}-1}^{\beta i}$, where $\beta$ runs through all non-zero elements of $\F_{p^{d+1}}$. This means that evaluation $\chi_\beta(x) = \omega^{\beta \log_\alpha(x)}$ requires the computation of a discrete log over the multiplicative group of the field. 

It is known that for non-trivial $\psi$ and $\chi$, the absolute value of the Gauss sum $G(\psi,\chi)$ evaluates to $|G(\psi,\chi)|=\sqrt{q}$, i.e., $G(\psi,\chi) = \sqrt{q} e^{i \theta}$, where $\theta \in [0,2\pi)$. The paper \cite{vDS:2002} established that $\theta$ can be approximated with precision $\varepsilon$ by a quantum algorithm in time $O(\frac{1}{\varepsilon} polylog( q))$. As we are overall only looking for a quantum algorithm that can solve the hidden shift problem over $\Z_N$ with bounded probability of success, it will be enough to approximate the diagonal elements of $\Delta$ with constant precision, i.e., we can use the van Dam/Seroussi algorithm to estimate $G(\psi,\chi)$. A minor complication is the fact that in \cite{vDS:2002} only the case of known character $\chi$ is considered, however, by making all steps of the algorithm conditioned on the character $\chi$ it can be easily seen that the transformation $\ket{\chi} \mapsto G(\chi,\psi)/\sqrt{q} \ket{\chi}$ can also be implemented coherently, i.e., on superposition of inputs $\chi$. The final step is to show how to relate $\chi(D)$ and $G(\chi,\psi)$. For this we make the restriction that $p=2$ so that our parameters always take the form $N=2^{d+1}-1$. We then obtain that
\begin{eqnarray*}
\chi(D) &=& \sum_{x : {\rm tr}(x)=0} \chi(x) = \sum_{x \in \F_q^\times}\chi(x) (1 + (-1)^{{\rm tr}(x)})=\sum_{x \in \F_q^\times}\chi(x) + \sum_{x \in \F_q^\times}\chi(x)(-1)^{{\rm tr}(x)}\\
&=& 
\sum_{x \in \F_q^\times}\chi(x)(-1)^{{\rm tr}(x)} = G(\psi,\chi),
\end{eqnarray*}
where $\psi$ denotes the additive character $\psi(x) := (-1)^{{\rm tr}(x)}$ of $\F_{2^n}$ and $\chi(x)$ denotes a multiplicative character of $\F_{2^n}^\times$. This argument establishes that we can approximate the operator $\Delta$ efficiently on a quantum computer with constant precision $\varepsilon$. 

The final two steps of the algorithm are easy to do: Step 5 is just another Fourier transform and Step 6 a measurement in the computational basis, both of which can be done efficiently. 

Step (iii): To construct the desired instances of the hidden subgroup problem from the hidden shift problem, we apply the results from Subsection \ref{sec:inj} and specialize them to the case of the Singer difference sets. We pick $m=2(d+1)+6$ random elements $v_1, \ldots, v_m \in \F_{q^{d+1}}$ and construct the hiding function $g_{v_1,\ldots, v_m}(x) := (g(x+v_1),\ldots ,g(x+v_m))$ which according to Theorem \ref{th:injectivize} is injective with probability greater than $1-\frac{1}{64}$. We then apply another standard construction \cite{Friedl2003,Kuperberg} which allows to turn an instance of an injective hidden shift problem into a hidden subgroup problem. Indeed, if $f, g: A \rightarrow S$ is an injective instance of a hidden shift problem with shift $s\in A$, then the corresponding hidden subgroup problem over $A \rtimes \Z_2$ is given by the hiding function $F((a,0)) := f(a)$ and $F((a,1)) := g(a)$, where $(a,t)$ is an encoding of the elements, i.e., $a\in A$ and $t\in \Z_2$. Conversely, if $F: A \rtimes \Z_2 \rightarrow S$ is a defining function of a hidden subgroup problem with hidden subgroup $H=\langle (a,1)\rangle$ of order $2$, then $f(x) := F(x,0)$ and $g(x) := F(x,1)$ defines a hidden shift problem over $A$. 

Overall, we established the claimed result of the existence of an efficient quantum algorithm to solve the hidden subgroup problem. The classical complexity of finding the shift $s$ from $\beta$ clearly is as least as hard as solving the discrete logarithm over a finite field. 
\qed 

\begin{corollary}\label{cor:instances} Let $N=2^n-1$, where $n\geq 2$, there there exist an expected number of $2^{n^2}$ instances of hidden subgroup problems over $D_N$ that can be solved efficiently on a quantum computer. 
\end{corollary}
\proof From the proof of Theorem \ref{th:dihedral} we see that in step (iii) for each random choice of $m=O(\log{|A|})$ elements, where $|A|=|\Z_{2^n-1}|=2^n-1$, we obtain a valid injectivization of the hidden shift function. There are an expected number of $O(|A|^m) = O((2^{\log_2(|A|)})^m) = O(2^{n^2})$ such functions. 
\qed

%
%

\section{Conclusions}
\label{conclusions}

We showed that the property of difference sets to give rise to functions with two level Fourier (power) spectrum which makes them useful for classical applications also allows to define hidden shift problems which can then be tackled on a quantum computer. While a solution to general hidden shift problems for arbitrary difference sets remains elusive, we showed that several interesting special cases can indeed be solved efficiently on a quantum computer. This includes the known cases of the Legendre symbol which we show to be an instantiation of our framework for the case of a Paley difference set. Furthermore, it includes the case of hidden bent functions which we show to be special cases of Hadamard difference sets. The case of Singer difference sets appears to be new and allows us to construct white-box instances of dihedral hidden subgroup problems that can be solved fully efficiently on a quantum computer, both in the quantum and in the classical parts of the algorithm. 

Open problems include whether these findings have any consequence for more general classes of instances of the dihedral hidden subgroup problem and the hidden subgroup problem in other semidirect products of a similar form. Other open problems include whether it is possible to solve the shifted difference set problem for projective planes which we mentioned cannot be solved by our main algorithm with better than exponentially small probability of success. One possible avenue for future research is to consider multi-register algorithms to tackle this problem. Another open problem is the case of hidden shift problems over abelian groups for functions that have approximately constant spectra, possibly with the exception of the zero frequency as in case of the functions arising from difference sets considered in this paper. 

\section*{Acknowledgments}
The author would like to thank Schloss Dagstuhl for hosting Seminar 15371, during which part of this research was carried out.

\bibliography{main}

\end{document}